\documentclass[a4paper,fleqn]{cas-dc}

\usepackage[square,sort,comma,numbers]{natbib}

\def\tsc#1{\csdef{#1}{\textsc{\lowercase{#1}}\xspace}}
\tsc{WGM}
\tsc{QE}
\tsc{EP}
\tsc{PMS}
\tsc{BEC}
\tsc{DE}

\begin{document}
\let\WriteBookmarks\relax
\def\floatpagepagefraction{1}
\def\textpagefraction{.001}
\let\printorcid\relax

\shorttitle{A generalized simplicial model and its application}

\shortauthors{Rongmei Yang et~al.}

\title [mode = title]{A generalized simplicial model and its application}                      



%
\author[1]{Rongmei Yang}[style=chinese]

\credit{Conceptualization, Methodology, Software, Writing - original draft, Writing - review \& editing}

\affiliation[1]{organization={Institute of Fundamental and Frontier Sciences, University of Electronic Science and Technology of China},
    city={Chengdu},
    postcode={610054}, 
    country={P. R. China}}
    
\affiliation[2]{organization={School of Cyber Science and Technology, University of Science and Technology of China},
    city={Hefei},
    postcode={230026}, 
    country={P. R. China}}
    
\affiliation[3]{organization={Yangtze Delta Region Institute (Huzhou), University of Electronic Science and Technology of China},
    city={Huzhou},
    postcode={313001}, 
    country={P. R. China}}

\author[1,3]{Fang Zhou}[style=chinese]

\cormark[1]

\ead{zervel3@std.uestc.edu.cn}

\credit{Conceptualization, Methodology, Software, Writing - original draft, Writing - review \& editing}

\author[1,3]{Bo Liu}[style=chinese]

\credit{Conceptualization, Methodology, Writing - review \& editing}

\author[2,1]{Linyuan L\"u}[style=chinese]
\cormark[1]
\ead{linyuan.lv@uestc.edu.cn}
\credit{Conceptualization, Funding acquisition, Project administration, Supervision, Validation, Writing - original draft, Writing - review \& editing}

\cortext[cor1]{Corresponding author}

\begin{abstract}
Higher-order structures, consisting of more than two individuals, provide a new perspective to reveal the missed non-trivial characteristics under pairwise networks. Prior works have researched various higher-order networks, but research for evaluating the effects of higher-order structures on network functions is still scarce. In this paper, we propose a framework to quantify the effects of higher-order structures (e.g., $2$-simplex) and vital functions of complex networks by comparing the original network with its simplicial model. We provide a simplicial model that can regulate the quantity of $2$-simplices and simultaneously fix the degree sequence. Although the algorithm is proposed to control the quantity of $2$-simplices, results indicate it can also indirectly control simplexes more than $2$-order. Experiments on spreading dynamics, pinning control, network robustness, and community detection have shown that regulating the quantity of $2$-simplices changes network performance significantly. In conclusion, the proposed framework is a general and effective tool for linking higher-order structures with network functions. It can be regarded as a reference object in other applications and can deepen our understanding of the correlation between micro-level network structures and global network functions.
\end{abstract}




\maketitle

\section{Introduction}
A complex network is a network-based description of a complex system. In the past decades, a large number of studies have explored the relationship between the structure and function of pairwise networks (i.e., networks consisting of nodes and edges)\cite{newman2003structure}. Structure and dynamics are the focus of much research, enabling a better understanding of networks' dynamical and functional behavior\cite{boccaletti2006complex}. Many studies have explored how networks with different structures affect dynamic behavior from the pairwise perspective\cite{chen2013identifying,pastor2002epidemic,bogua2003epidemic,barthelemy2005dynamical}. However, empirical observations indicate the existence of complex contagions in reality\cite{centola2007complex,centola2010spread,karsai2014complex}. For example, some individuals in a community adopt a new product, and an individual is convinced to adopt the product by the adopted individuals of the community.  To systematically describe the interaction involving multiple individuals, we need more complex associations between network constituents. Therefore, investigating higher-order networks (i.e., networks representing many-body interactions) has become a prominent focus in network structure and dynamics\cite{bianconi2021higher}.

Higher-order networks provide a new perspective to describe networks such as social networks\cite{alvarez2021evolutionary}, biological networks\cite{sanchez2019high}, and brain networks\cite{sizemore2018cliques}, which are beyond the traditional pairwise connectivity framework. There are mainly two methodologies to describe higher-order networks: simplicial networks\cite{spanier1989algebraic} and hypergraphs\cite{iacopini2019simplicial}.  In this paper, we focus on simplicial networks, which provide a powerful tool to describe the higher-order structure of networks. Simplicial networks represent the rich higher-order structure in network data through simplices. The concept of the simplex is derived from topological algebra\cite{spanier1989algebraic}, where a node is defined as a $0$-simplex, an edge as a $1$-simplex, and a triangle as a $2$-simplex, etc.

Recently, many studies have indicated that higher-order structures can enhance the comprehension of dynamic behaviors and functional implementations within complex systems\cite{battiston2020networks,torres2021and,battiston2021physics,shi2019totally}. Reimann et al. analyzed the intricate higher-order structures, such as cliques and cavities in neural networks, revealing the brain processes stimuli and complex functions\cite{reimann2017cliques}. Sizemore et al. studied the presence of large cliques and cavities in mesoscale network structures that play a unique role in controlling brain function\cite{sizemore2018cliques}. Shi et al. discovered that totally homogeneous networks are networks with the best possible synchronizability\cite{shi2013searching,shi2019totally}. Iacopini et al. proposed a higher-order spreading model, which induced the discontinuous transition when the higher-order interaction is considered\cite{iacopini2019simplicial}. Torres and Bianconi discussed higher-order diffusion's spectrum and return probability\cite{torres2020simplicial}. Unlike the classical random walk, Schaub et al. and Mukherjee et al. studied the random walk in higher-order structure\cite{schaub2020random, mukherjee2016random}. 

Higher-order structures are widespread in the real world, and using higher-order structures to model network systems may discover new non-trivial properties missed in pairwise perspective. To better understand the significance of higher-order structures, scholars paid attention to controlling the higher-order structures and developing models to construct networks that contain higher-order structures. Courtney and Bianconi designed a configuration model named the NGF model that keeps the generalized degree sequence fixed and thus captures the fundamental properties of one, two, three, or more linked nodes\cite{courtney2016generalized}. Kovalenko et al. generated networks with scale-free $2$-simplices and generalized the NGF model by proposing a combination model of preferential and nonpreferential rules\cite{kovalenko2021growing}. Bobrowski and Krioukov reviewed the typical random simplicial complexes models\cite{bobrowski2022random}. Costa and Farber controlled the number of different order simplex by a group of parameters\cite{costa2016random}. Yen designed a configuration model that guarantees the degree sequence and fact size distribution\cite{yen2021construction}.


Although a large quantity of studies focuses on the higher-order perspective, the work on how higher-order structure influences network functions systematically is still out of the way. Therefore, a framework connecting structure and function through higher-order network modeling is indispensable. 
In this paper, we propose a simplicial model that enables simultaneous control of the quantity of $2$-simplices and guarantees a fixed degree sequence. The model constructs simplicial networks by abstracting pairwise networks into simplices, thus providing access to a large number of real-world networks. Classical applications, such as spreading dynamics, pinning control, network robustness, and community detection, are employed to demonstrate that the proposed model offers an efficient framework for quantifying the impact of higher-order structures on network functions.





\section{Methods \& Data}
\subsection{Simplices and simplicial networks}
\begin{figure*}[htbp]
    \centering
    \includegraphics[width=0.9\textwidth]{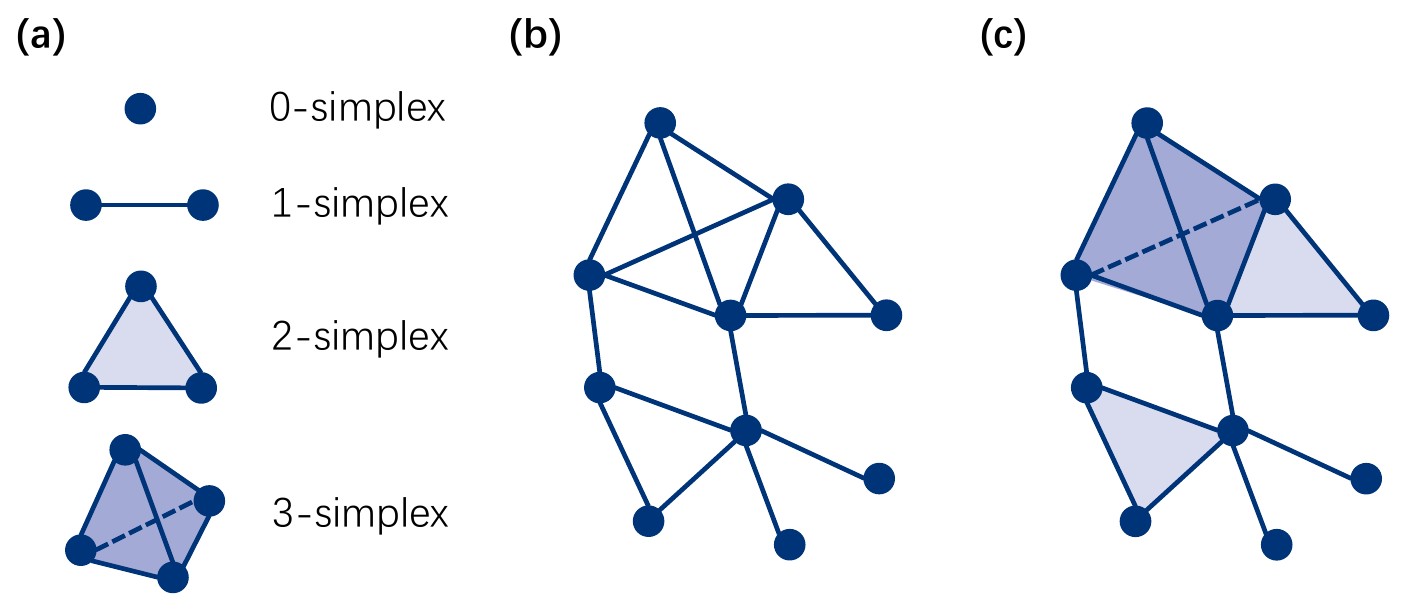}
    \caption{{\bf The translation from a pairwise network to a simplicial network.} {\bf (a)} The $m$-simplex, which forms simplicial networks. {\bf (b)} The example of pairwise network. {\bf (c)} The clique complex is constructed from the pairwise network in {\bf (b)}, by considering each clique as a simplex. }

    \label{fig: clique_complex.pdf}
\end{figure*}

Higher-order structures can effectively describe simplicial networks.
To ensure the applicability of our proposed model, we generate simplices from pairwise networks and consider them as a simplicial network.  The simplices represent interactions involving two or more nodes and are defined from algebraic topology\cite{spanier1989algebraic}. For a network $G$, a complete subgraph with $m+1$ nodes is called an $m$-simplex,  e.g., a node is a $0$-simplex, an edge is a $1$-simplex, a triangle is a $2$-simplex, and a tetrahedron is a $3$-simplex. The quantity of $m$-simplices in the network is denoted as $S_m$. Simplicial complexes are formed by a set of simplexes and are used to describe simplicial networks. We first obtain the clique complex of the network by finding all cliques in the network and each clique is considered as a simplex in the simplicial network. Fig. \ref{fig: clique_complex.pdf} shows an example of related concepts including simplexes, pairwise networks, and clique complexes.

\subsection{Simplicial algorithm}

 To control the quantity of $2$-simplices, we proposed an iterative algorithm that generates or dismantles $2$-simplex by rewiring the edges. Specifically, given a network, at each step, the algorithm will generate (dismantle) at least one $2$-simplex by rewiring the edges, and obey the following three rules: (\romannumeral1) keep the network connected; (\romannumeral2) without multiple edges and self-loops; (\romannumeral3) fix the degree sequence of the original network. The algorithm will stop when it reaches the given reselection times $R$. We denote the total successful rewiring times as $n$, and if the successful rewiring times reach $n$, we denote that the generated network has generated or dismantled $100\%$ $2$-simplex. During the iteration, the generated network will be saved for each of $0.1 n$ of successful rewire, and all the results are averaged by $10$ realizations.
 
 \begin{figure}
    \centering
    \includegraphics[width=1\linewidth]{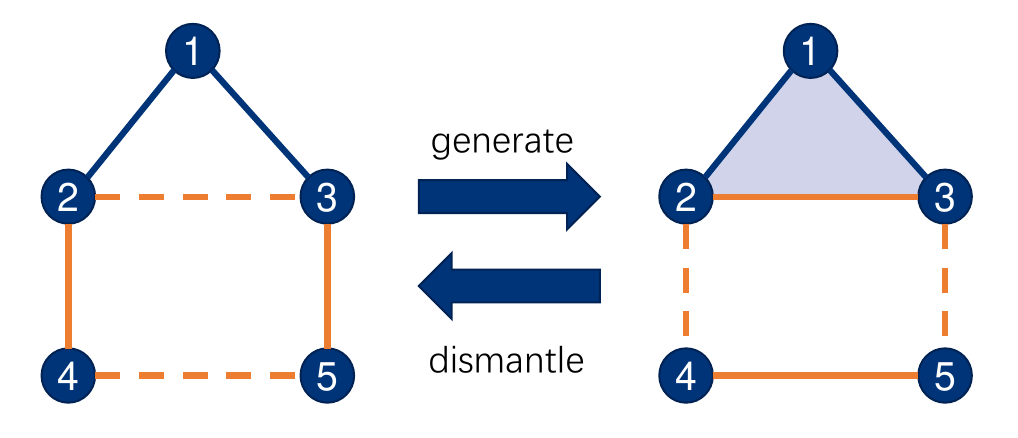}
    \caption{{\bf The process to generate and dismantle $2$-simplex.}}
    \label{fig:rewire.pdf}
\end{figure}

 We take Fig. \ref{fig:rewire.pdf} to explain how to regulate $2$-simplex and the process of how to generate $2$-simplices is described as follows:
 
\begin{enumerate}
\item Randomly select a node $v_{1}$ satisfying degree $d_{v_1} \ge 2$ with uniform probability.
\item Randomly choose nodes $v_2$, $v_3$ from node $v_1$’s neighbors, satisfying degree $d_{v_2}, d_{v_3} \ge 2$  and nodes $v_2$, $v_3$ are not connected. If not, turn to step $1$.
\item Choose node $v_4$ ($v_5$) from the neighbors of $v_2$ ($v_3$), and meet the following conditions:
(a) $v_{4}$ ($v_5$) is not $v_{1}$;
(b) $v_{4}$ ($v_5$) has no common neighbors with $v_{2}$ ($v_3$).
Otherwise, go back to step $2$.
\item Remove edges $(v_2, v_4)$, $(v_3, v_5)$, and connect edges $(v_2, v_3)$, $(v_4, v_5)$. If the network is connected, we generate at least one $2$-simplex and call it a successful rewiring. Otherwise, we restore the connection and start again from step $1$.

\item If the reselection of node $v_1$ reaches the $R$ times, stop the rewiring process.
\end{enumerate}

The algorithm to dismantle $2$-simplex is described as follows:
\begin{enumerate}
\item Randomly select a $2$-simplex $(v_{1},v_{2},v_{3})$ with uniform probability.
\item Select an edge $(v_4, v_5)$ randomly, and nodes $v_4$, $v_5$ are not connected with any node of the $2$-simplex selected in step $1$.
\item Remove edges $(v_2, v_3)$, $(v_4, v_5)$, and connect edges $(v_2, v_4)$, $(v_3, v_5)$. If the network is connected, we dismantle at least one $2$-simplex and call it a successful rewiring. Otherwise, we restore the connection and start again from step $1$.

\item If the reselection reaches the $R$ times, stop the rewiring process.
\end{enumerate}

\subsection{Empirical data}
In this paper, we use four empirical networks to support our experiments, including the neural network of \emph{Caenorhabditis elegans} (\emph{C.elegans})\cite{C.elegans}, the Wikipedia voting data from the inception of Wikipedia till January 2008 (\emph{Wiki-Vote})\cite{wikivote}, the E-mail communication network between college students (\emph{Email})\cite{email}, and the \emph{E. coli} metabolic network (\emph{Metabolic})\cite{metabolic}. The basic structural features of the four networks are shown in Table \ref{fig: network size}.

\begin{table}[H]
\centering
\caption{The basic statistics of the four empirical networks. $N$ is the size of the network. $M$ is the number of edges. $\left \langle k \right \rangle$ is the average degree. $c$ is the clustering coefficient. $\left \langle L \right \rangle$ is the average path length. $r$ is the degree assortativity coefficient. $S_2$ is the quantity of $2$-simplices. }
\resizebox{\linewidth}{!}{
\begin{tabular}{cccccccc} 
\hline
& $N$ & $M$ & $\left \langle k \right \rangle$ & $c$ & $\left \langle L \right \rangle$ & $r$ & $S_2$  \\ 
\hline
\emph{C.elegans} & 297  & 2148  & 14.465 & 0.292 & 2.447 & -0.163 &  3241 \\
\emph{Wiki-vote} & 889  & 2914  & 6.555  & 0.152 & 4.091 & -0.029 &  2119 \\
\emph{Email}     & 1133 & 5451  & 9.622  & 0.220 & 3.602 &  0.078 &  5343 \\
\emph{Metabolic} & 1039 & 4741  & 9.126 & 0.377  & 2.474 & -0.250 &  6876 \\

\hline
\end{tabular}}
\label{fig: network size}
\end{table}

\section{Results}
\subsection{Quantity of $m$-simplices}
\begin{figure*}[htbp]
    \centering
    \includegraphics[width=1\textwidth]{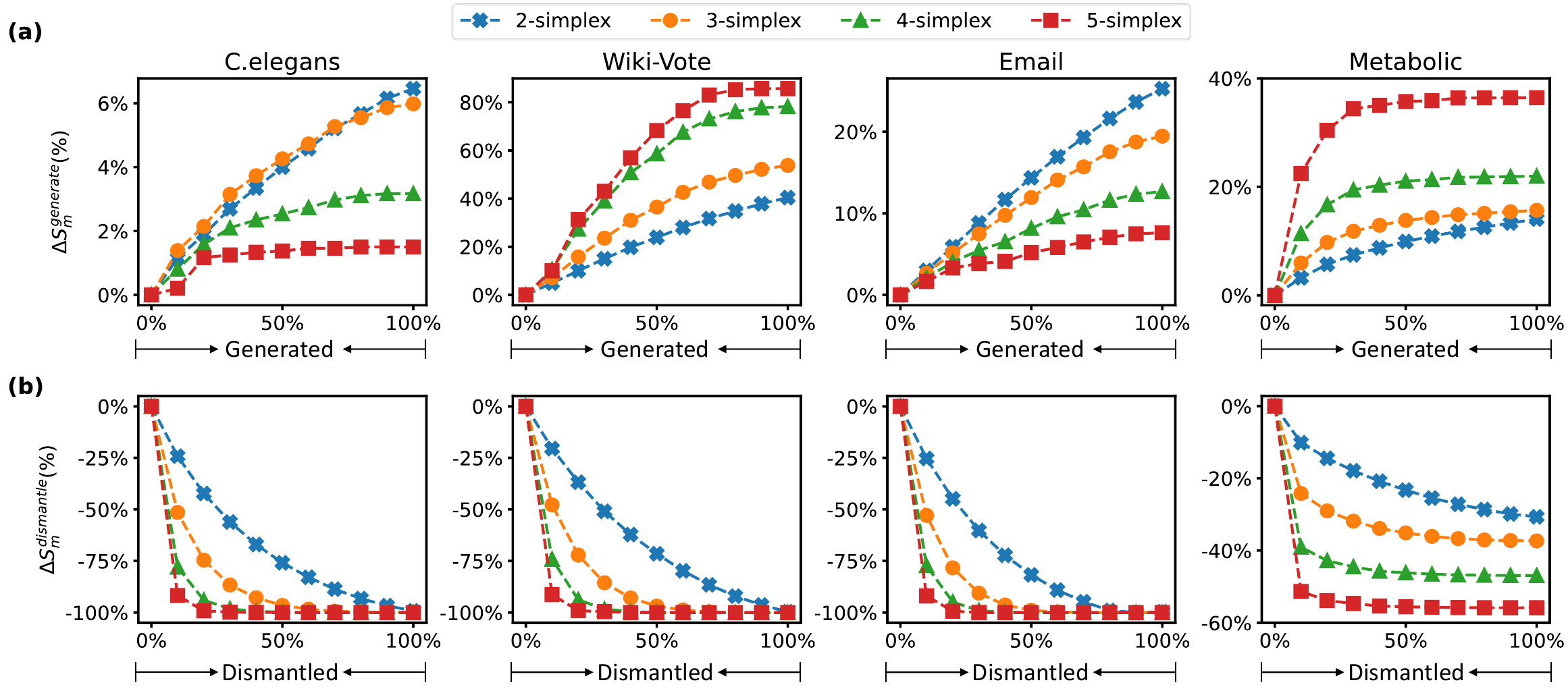}
    \caption{{\bf Correlation between the fraction of rewiring times and quantity of $m$-simplex.} Using the proposed simplicial model, we generate or dismantle $2$-simplex for a given dataset. The $x$-axis denotes the ratio of generated or dismantled $2$-simplex of the generated network. The $y$-axis measures the changing ratio of $S_{m}$ between the network generated by the simplicial model and the original network.}
    \label{fig:change of k-clique}
\end{figure*}

Fig.\ref{fig:change of k-clique} presents the $m$-simplex ($m=2, 3, 4, 5$) of four networks generated by the simplicial model. The first row is the result of the generated $m$-simplex, and the second row is the result of the dismantled $m$-simplex. The $x$-axis is the fraction of rewiring times of a given network, and the $y$-axis is the fraction of simplex that has increased or decreased compared with the original network. The results illustrate the quantity of $2$-simplices is positively correlated with higher-order simplex ($m>2$), and the change rate of the simplex is approximately logarithmic. Restricted by degree distribution, the quantity of generated simplex is limited. The correlation between the rewiring times and the Betti number\cite{shi2021computing} (the number of cavities, which is another significant higher-order structure) is shown in Supplementary Fig. A1.

\subsection{Spreading dynamics}

\begin{figure*}[htbp]
    \centering
    \includegraphics[width=0.9\textwidth]{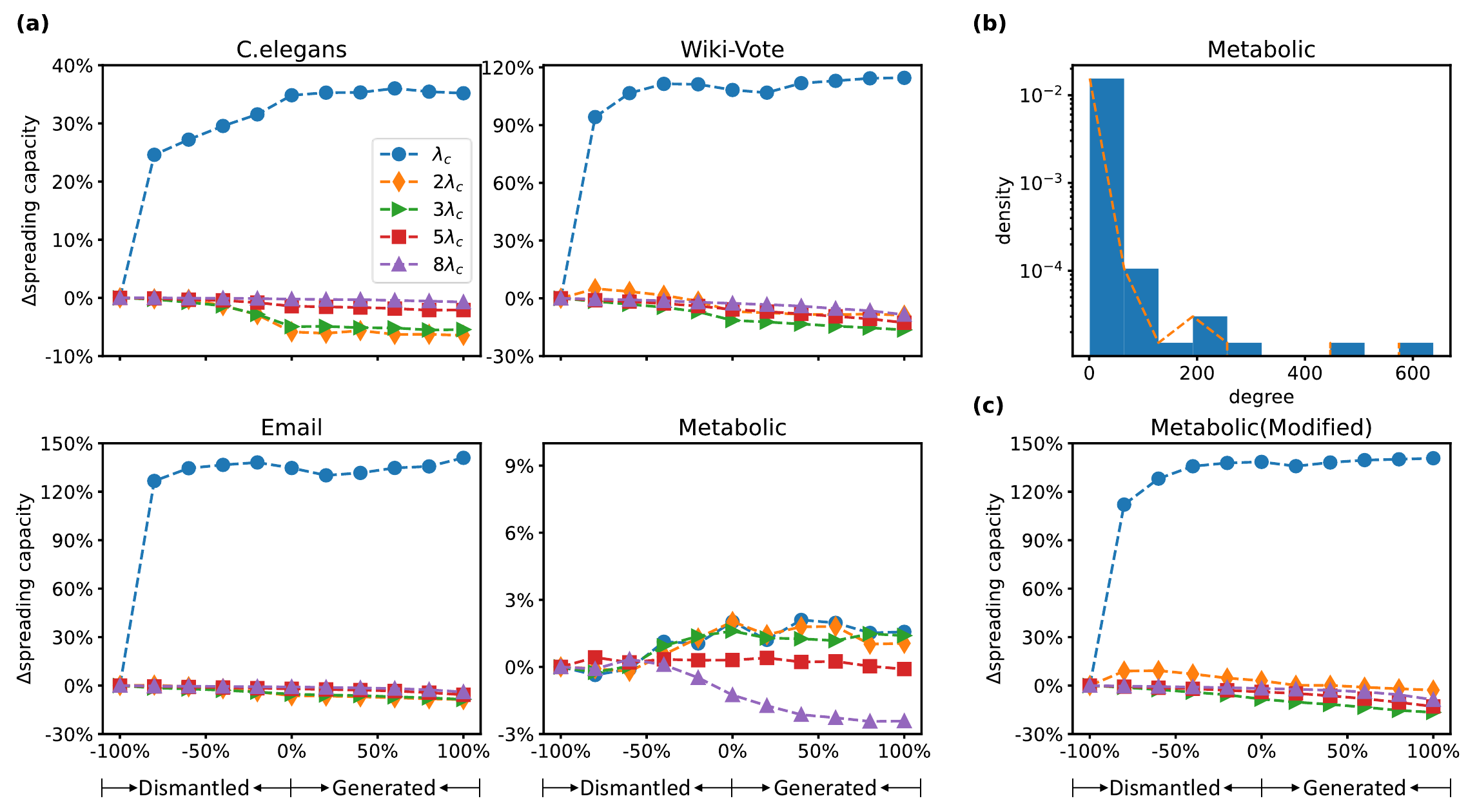}
    \caption{{\bf The influence of $2$-simplex on spreading dynamics.} {\bf (a)} The spreading capacity of top individuals under different infection rates. The $x$-axis denotes the ratio of generated or dismantled $2$-simplex of the generated network. The $y$-axis denotes the improved spreading capacity compared with the top individuals in the network with the minimum quantity of $2$-simplices. {\bf (b)} The degree distribution of the original \emph{Metabolic} network. {\bf (c)} The performance of \emph{Metabolic} network after removing top $1\%$ degree nodes.}
    \label{fig:spread capacity}
\end{figure*}

To consider the influence of $2$-simplex on the spreading dynamics, we construct networks with different quantities of $2$-simplices with the proposed simplicial model and employ the Susceptible-Infected-Recovered (SIR) model to simulate the spreading process. There are two parameters $\beta$ and $\gamma$ in the SIR model, where $\beta$ is the probability an infected individual infects its neighbor, and $\gamma$ is the probability an infected individual recovers to the recovered state. In this paper, we fix the $\gamma=1$ and choose several groups of infection rate $\beta$, including $\beta=\lambda_c$, $2\lambda_c$, $3\lambda_c$, $5\lambda_c$, and $8\lambda_c$.  Here the $\lambda_c$ is the epidemic threshold of the SIR model\cite{shu2015numerical,boguna2002epidemic}, which is correlated with the network structure, and the definition is as follows:

\begin{align}
\lambda_c & = \frac{\left \langle k \right \rangle }{\left \langle k^2 \right \rangle - \left \langle k \right \rangle}, 
\end{align}
where $\left \langle k \right \rangle$ and $\left \langle k^2 \right \rangle$ are the average degree and the average squared degree of the network. Since the networks generated by the simplicial model have the same degree distribution, their epidemic threshold is the same (the epidemic threshold for each network is shown in Supplementary Table A2. For a generated network, we choose the top $1\%$ of individuals ranked by degree $k$ and $S_2$, respectively, as seeds to simulate the spreading process. The dynamic process will stop when no individual is in the infected state, and we denote the 
fraction of recovered individuals as the spreading capacity of seeds.
Given a dataset, a total of $10$ networks generated by the simplicial model, 
we observe how the increase of $2$-simplex influences the spreading capacity of seeds. After we get the spreading capacity of the seeds of a given network, we minus the spreading capacity of the seeds of the network with the minimum quantity of $2$-simplices.

The result of seeds ranked by $S_2$ is presented in Fig.~\ref{fig:spread capacity}~(a) (the result ranked by degree shows a consistent trend, and see Supplementary Fig. A4 for details), which reveals two insights. Firstly, as the quantity of $2$-simplices increases, the spreading capacity of top individuals improves tremendously at the epidemic threshold, and compared with the network with the minimum quantity of $2$-simplices, the spreading capacity of the network with maximum quantity improves from $30\%$ to more than $100\%$ on the three datasets (except for \emph{Metabolic}).  Secondly, when the infection rate increases from $2\lambda_c$ to $8\lambda_c$, the spreading capacity even decreases slightly, indicating an increase in the quantity of $2$-simplices will restrain the spreading, which is consistent with the previous research\cite{chen2013identifying}.

To distinguish why the spreading capacity of \emph{Metabolic} at the epidemic threshold performs differently from the other three datasets, we focus on the network properties of the four datasets and find that compared with the other three datasets, the networks generated by the \emph{Metabolic} dataset have approximately the same degree assortativity (see Supplementary Fig. A2 for details). To measure whether the degree assortativity is the key factor that causes the above phenomenon, we focus on the degree distribution of \emph{Metabolic} and find it's different from the other three networks. Specifically, the degree of a small fraction of individuals is abnormally large in \emph{Metabolic} (see Fig.~\ref{fig:spread capacity}~(b)). After removing these individuals and linked edges, we rerun the simplicial model and found that the increase of the $2$-simplex will increase the degree assortativity (see Supplementary Fig. A2 for details). Finally, the four empirical datasets show consistent performance (Fig.~\ref{fig:spread capacity}~(c)). To make the result more robust, we test two synthetic datasets,  scale-free and random datasets, which also show consistent performance (see Supplementary Figs. A3 and A4 for details). 

In summary, we find an increase in the quantity of $2$-simplices will restrain the spreading when the infection rate exceeds the epidemic threshold; on the contrary, when the infection rate is at the epidemic threshold, it will prompt the spreading process. This is a new finding about how higher-order structure influences the spreading dynamics.

\subsection{Pinning control} 

\begin{figure*}[htbp]
    \centering
    \includegraphics[width=1\textwidth]{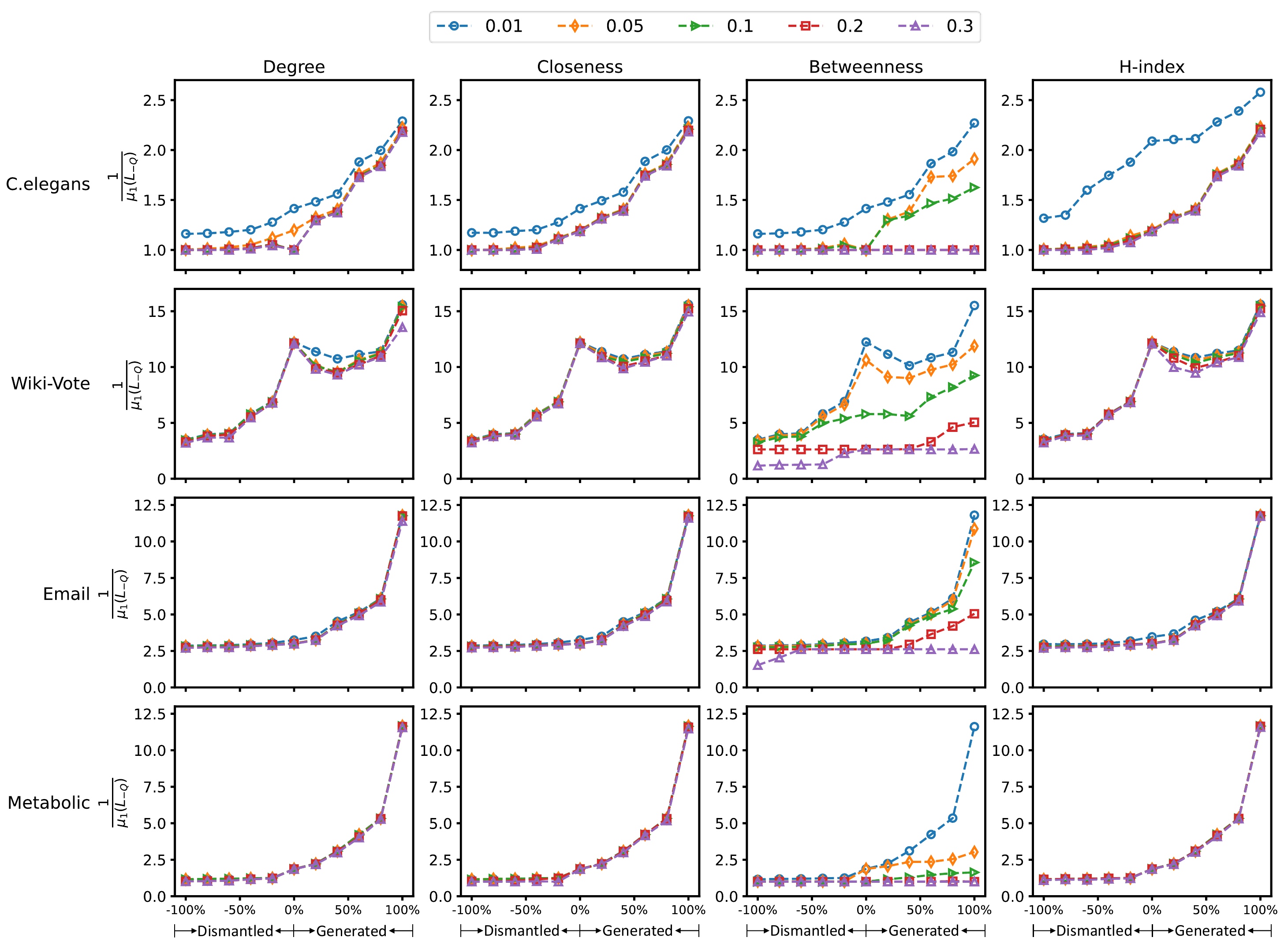}
    \caption{{\bf The influence of $2$-simplex on pinning control effect.} For the original network and networks generated by the simplicial model, we pin nodes by degree, closeness, betweenness, and h-index, respectively. The $x$-axis denotes the ratio of generated or dismantled $2$-simplex of the generated network. Each curve in the sub-figure denotes the result that is pinned by different fractions of nodes. The $y$-axis denotes the synchronizability after pinning the corresponding fraction of nodes.}
    \label{fig:pinning}
\end{figure*}

The concept of pinning control was first proposed in the field of partial differential equations\cite{grigoriev1997pinning}, then it was introduced into complex networks fields to describe a feedback control action exerted on a fraction of nodes\cite{wang2002synchronization}. Specifically, pinning control is used to measure the importance of the nodes by pinning these nodes in a synchronizing process\cite{wang2002pinning, li2004pinning, fan2021characterizing}. Given a network $G(V, E)$ with $N$ linearly and diffusively coupled nodes, we denote the interacting dynamic is 
\begin{equation}
    \dot x_i = f(x_i) + \sigma\sum_{j=1}^NL_{ij}\Gamma (x_j) +U_i (x_i,...,x_N).
\end{equation}
Here the $x_i$ is the state vector of node $i$, the function $f(x_i)$ describes the self-dynamics of node $i$, $\sigma$ is the coupling strength, $U_i$ is the controller applied at node $i$, and the positive semi-definite matrix $\Gamma$ is the inner coupling matrix. The Laplacian matrix $L=[L_{ij}]_{N \times N}$ of $G$ is defined as follows: if $(i, j)\in E$, $L_{ij}=-1$ holds; if $(i, j)\not\in E$, $L_{ij}=0$ holds; if $i=j$, $L_{ii}=k_i$ holds; and $k_i$ is the degree of node $i$.

Pinning control is a process that drives the system from any initial state to the target state in a finite time by pinning some selected nodes. The performance of pinning control is measured by the synchronizability of the pinned network, and represented by the reciprocal of the smallest nonzero eigenvalue of the principal submatrix, which is obtained by deleting the corresponding rows and columns of the pinned nodes of the adjacent matrix. The smaller the smallest nonzero eigenvalue is, the better the pinning control effect is. 

 In this section, we discuss how a $2$-simplex influences the pinning control effect. To achieve the goal, for a given network generated by our simplicial model, we pin at most the top $30\%$ nodes ranked by a centrality metric, including degree,  closeness, betweenness, and H-index. Fig.\ref{fig:pinning} shows the pinning control results of the four datasets, where each column denotes the result obtained by a centrality metric. The $x$-axis indicates which networks (generated or dismantled by the model) are represented. And the $y$-axis is the pinning control effect. The smaller the value of the $y$-axis is, the better the performance is. To show a clear performance on how $2$-simplices influence the pinning control effect, We only give the results of pinned nodes in the fraction of $0.01$, $0.05$, $0.1$, $0.2$, and $0.3$. According to Fig. \ref{fig:pinning}, networks with less number of $2$-simplices are easier to pin, and the pinning control effect is better, which indicates the quantity of $2$-simplices will restrain the pinning control effect. In the \emph{Wiki-Vote} dataset, we find a different increasing trend compared with the other $3$ datasets, that is for the original network, the pinning control effect is worse than the neighboring networks. However, the corresponding quantities of $2$-simplices in original and generated \emph{Wiki-Vote} networks are strictly monotonically increasing with the increase of iteration times. It indicates except for the $2$-simplex, there are other structure features that can also influence the pinning control effect. To further explore which structure features cause the above phenomenon, we employ three types of null models\cite{xu2020quantifying} to randomize the original \emph{Wiki-Vote} dataset (null models can control networks' different order structural features), then apply the simplicial model to the randomized datasets and proceed to the pinning control experiments. The result is displayed in supplementary Fig. A7. We find the original \emph{Wiki-Vote}'s pinning control effect is similar to its $2.5$-order and $3$-order null models, but different from its $2$-order null model. Based on the previous research \cite{hong2004factors,nishikawa2003heterogeneity,comellas2007synchronizability}, we find the network's betweenness value (maximal betweenness of nodes) will affect the pinning control effect as well. Specifically, the smaller the betweenness is, the better the pinning control effect is. 
Therefore, we present the betweenness value of the original network and its null models in Fig. A6, and find that compared with the other three datasets, the original \emph{Wiki-Vote} dataset gets a high betweenness value, but its neighboring rewired networks' betweenness value is low, which finally leads to a different pinning control effect for the \emph{Wiki-Vote} dataset.



\subsection{Network robustness}

\begin{figure*}[htbp]
    \centering
    \includegraphics[width=0.8\textwidth]{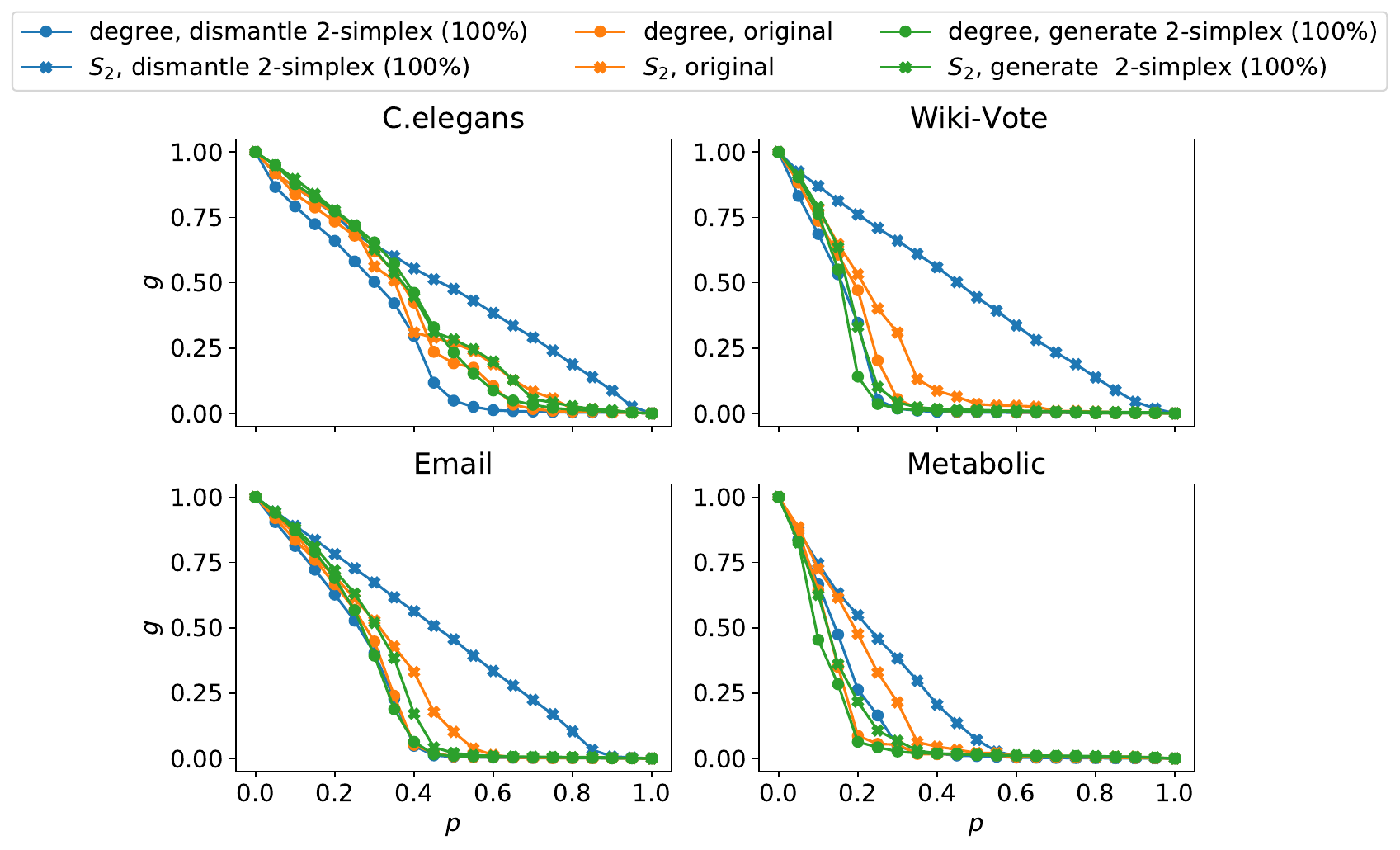}
    \caption{{\bf The impact of $2$-simplex on network robustness.} We attack nodes by their degree and quantity of $2$-simplices. Once the node is attacked, the node and its edges will be removed. The $x$-axis denotes the percentage of attacked nodes and the $y$-axis denotes the fraction of the size of the largest component after the attack. }
    \label{fig:percolation 2 index}
\end{figure*}

The network attack is a typical tool to measure the robustness of a network. Specifically,  if a node is attacked, then the node itself and its edges will be removed from the whole network. The attack is divided into random attack and targeted attack, with the former attacking the nodes randomly and the latter attacking the nodes by their ranking importance. The performance of an attack is usually measured by the size of the largest component after the attack. 

In our experiments, we choose to attack the network by two ranking methods: nodes ranked by degree and quantity of $2$-simplices $S_2$. The details of the attack process are as follows. Given a network $G$, we choose a node to attack at each step, removing the node itself and its edges. Then we calculate and record the size of the largest connected component of the network. The attack process is repeated until no node is left in the network. The network's robustness increases with the increasing size of the largest connected component, and the larger the largest connected component, the greater the network's robustness. 

 To reveal how $2$-simplex influences the network robustness under the perspective of the network attack,  for a given dataset, we only present the results of the original network and $4$ networks generated by our simplicial model: dismantling $50$\% and $100\%$ $2$-simplex,  generating $50\%$ and $100\%$ $2$-simplex.
Fig. \ref{fig:percolation 2 index} presents the results of the network attack effect ranked by degree and quantity of $2$-simplices. Two main conclusions can be drawn from the figure: 
(\romannumeral1) When we attack nodes by the ranking of $S_2$, the increase of $2$-simplex makes the network collapse faster. (\romannumeral2) For a certain network with the same quantity of $2$-simplices, it disintegrates faster when attacked by degree. Furthermore, we measure how $2$-simplex influences the robustness of scale-free and random networks, which keep consistent with the $4$ empirical datasets. Detailed results of scale-free and random networks are shown in Supplementary Fig. A5.

\subsection{Community detection}

\begin{figure*}[htbp]
    \centering
    \includegraphics[width=0.8\textwidth]{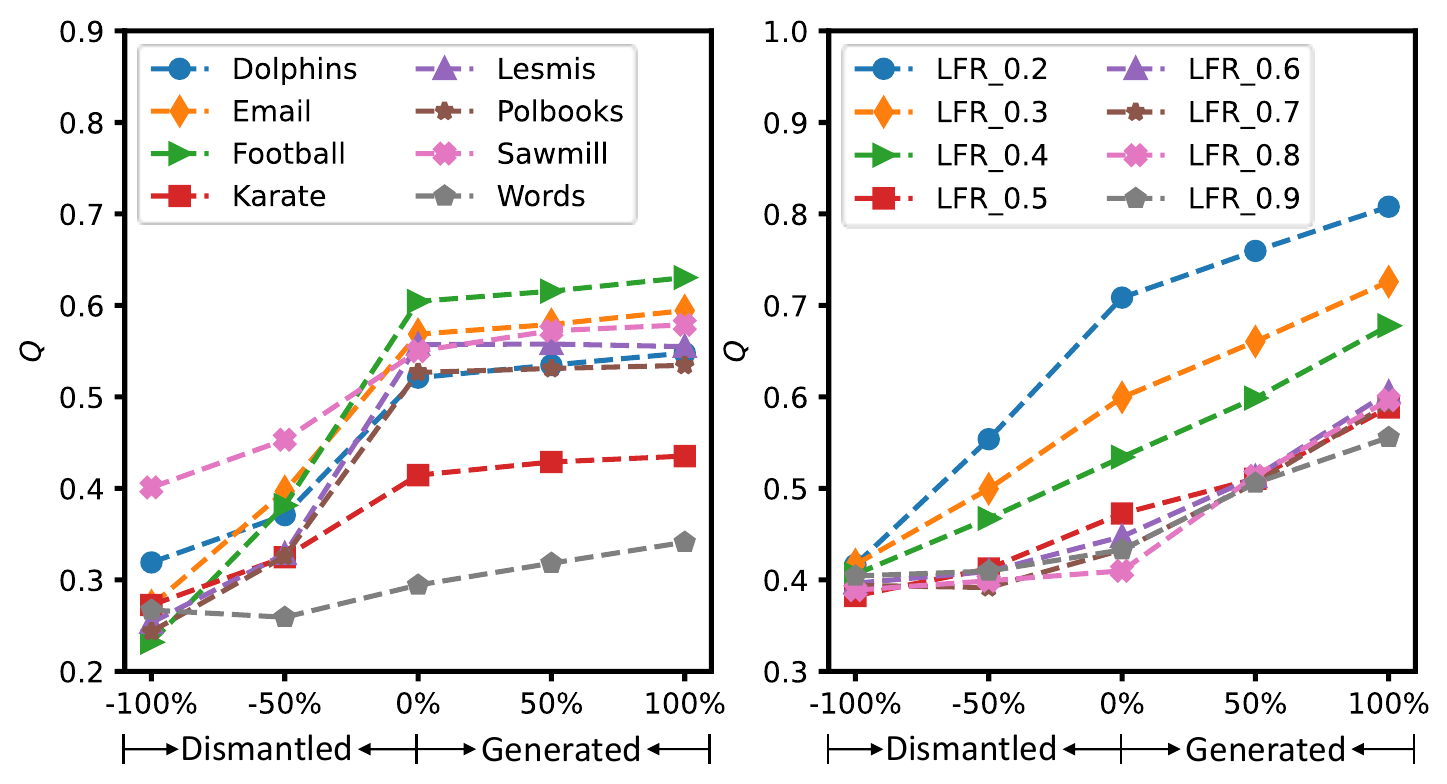}
    \caption{{\bf The impact of $2$-simplex on community detection in empirical networks and LFR networks.} The $x$-axis denotes different networks, from the minimum quantity ($-100\%$) of $2$-simplices to maximum quantity ($100\%$) of $2$-simplices. The $y$-axis denotes the modularity $Q$.}
    \label{fig:community}
\end{figure*}

To evaluate how the quantity of $2$-simplices influences the evolving of community structure, eight empirical networks are used in our experiment, which are \emph{Dolphins}\cite{Dolphin}, \emph{Email}\cite{email}, \emph{Football}\cite{football}, \emph{Karate}\cite{zachary1977information}, \emph{Lesmis}\cite{knuth1993stanford}, \emph{Polbooks}\cite{polbooks}, \emph{Sawmill}\cite{michael1997modeling} and \emph{Words}\cite{newman2006finding}. In addition, LFR network\cite{lfrbenchmark} is often used as the benchmark for community detection. We construct eight LFR networks with parameters $n=100, \langle k \rangle=5, k_{max}=10, c_{min}=5, c_{max}=10, \lambda=2, \beta=1$, and $\mu=0.1-0.8$. The metric modularity $Q$\cite{modularity,louvain} is usually used to measure the quality of community detection and is defined as

\begin{align}
Q=\frac{1}{2 M} \sum_{i j}\left(a_{i j}-\frac{k_i k_j}{2M}\right) \delta\left(c_i, c_j\right),
\end{align}
where $A=\left ( a_{ij} \right ) $ is the adjacency matrix of a given network, $k_i$ is the degree of node $i$, and $c_i$ is the community to which node $i$ is assigned. $\delta=1$ if the two nodes belong to the same community, otherwise $\delta=0$.

Fig.\ref{fig:community} shows community detection results using the Louvain algorithm\cite{louvain}. Here the $x$-axis denotes original networks and networks generated by our simplicial model (i.e., networks generated $50\%$ and $100\%$ $2$-simplices, networks dismantled $50\%$ and $100\%$ $2$-simplices), and the $y$-axis denotes the modularity $Q$. According to the figure, we find the increase of the quantity of $2$-simplices will improve the value of modularity $Q$, indicating that the community structure of networks becomes more significant as the quantity of $2$-simplices increases. Therefore, our study is helpful in improving the diversity of network community structure and contributes to the performance evaluation of community detection algorithms.

\section{Discussion}
High-order perspective provides an innovative tool to research the effect of higher-order structures on network functions and illustrates new phenomena missed under pairwise contagion\cite{iacopini2019simplicial}.
Our paper proposes a simplicial model that can control the quantity of $2$-simplices and fix the degree sequence simultaneously. Further, we investigate the effect of $2$-simplex on network functions, i.e.,  spreading dynamics, pinning control, network robustness, and community detection.

Previous works revealed that network clustering will influence the spreading dynamics\cite{chen2013identifying, holme2016temporal}, which is highly correlated with the quantity of $2$-simplices. Under the higher-order perspective, we find that increasing the quantity of $2$-simplices will suppress the spreading effect when the infection rate is larger than the epidemic threshold, which is consistent with former research\cite{chen2013identifying}.
However, the spreading effect improves tremendously with the increasing number of $2$-simplices when the infection rate equals the epidemic threshold, and further exploration finds that degree assortativity is the key point in causing such a phenomenon. Specifically, the increase of $2$-simplex will make the degree assortativity higher (sufficient but not necessary) and finally improve the spreading effect at the epidemic threshold, such that the result is consistent with the work\cite{ge2014epidemic}. The finding can be applied to designing strategy to control the dynamics, e.g., we can dismantle $2$-simplex to normal edges to restrain epidemic spreading when the infection rate is low. However, why the variation of degree assortativity can influence the spreading effect of individuals at the epidemic threshold needs theoretical interpretation. By applying the model to pinning control, we find that decreasing the quantity of $2$-simplices enables the whole network to update to the target state more quickly. Theoretically, if a network has a low quantity of $2$-simplices after the pinned nodes are fixed, the corresponding principal submatrix's smallest nonzero eigenvalue will be small. This indicates the decrease of $2$-simplex will prompt the synchronization process, and a strong local structure will hinder the synchronization. For the network robustness section, experiments show that the increase of $2$-simplex accelerates the collapse of networks, i.e., the size of the giant component will reduce quickly. 
The result suggests that the increase of $2$-simplex is harmful to network stability, and a compact local structure makes the whole network more fragile to external attacks. 
In terms of community detection, as the number of $2$-simplices increases,  the performance of the community detection algorithm improves. Such a result is obvious because increasing the number of $2$-simplices gives the entire network a more transparent local structure, which makes the community structure divided by the Louvain algorithm more distinct, and ultimately, improves the value of modularity.

Recently, a number of works have contributed to the simplicial models\cite{zuev2015exponential, courtney2016generalized, costa2016random, kovalenko2021growing, bobrowski2022random, yen2021construction}. Tzu-Chi Yen\cite{yen2021construction} proposed a model to generate a simplicial complex and keep the degree and fact sequence fixed simultaneously. However, this model requires high time complexity, and the generated networks are totally constituted by simplicial complexes. It is not suitable to analyze the empirical networks, which include not only simplices more than $2$ order but also pair-wise structure. Different from that model, the network generated by our simplicial model includes higher-order simplices and pair-wised structures, and our model is efficient as well. Further, except for the above applications, the model can also be applied to other fields, e.g., to explore how simplices influence the higher-order spreading effect. Our work provides an effective simplicial model to reveal what role the simplex plays in networks, and how the micro-level structure influences global functions. With the help of the simplicial model, we can research the higher-order interaction more deeply. Although our model doesn’t consider the simplex of more than $2$ orders, the result shows it can also indirectly control the quantity of higher-order simplex ($3$, $4$ orders, etc.).

\section*{Data \& Code Availability}
The datasets and codes used in this work are available at: https://github.com/meanzzzz/Generalized-simplicial-model.

\section*{Acknowledgements}
The authors acknowledge the STI 2030—Major Projects (Grant No. 2022ZD0211400), the National Natural Science Foundation of China (Grant No. T2293771), the Sichuan Science and Technology Program (Grant No. 2023NSFSC1919), and the New Cornerstone Science Foundation through the XPLORER PRIZE.


\printcredits

\bibliographystyle{unsrt}

\bibliography{reference}

\begin{thebibliography}{10}

\bibitem{newman2003structure}
Mark~EJ Newman.
\newblock The structure and function of complex networks.
\newblock {\em SIAM Review}, 45(2):167--256, 2003.

\bibitem{boccaletti2006complex}
Stefano Boccaletti, Vito Latora, Yamir Moreno, Martin Chavez, and D-U Hwang.
\newblock Complex networks: structure and dynamics.
\newblock {\em Physics Reports}, 424(4-5):175--308, 2006.

\bibitem{chen2013identifying}
Duan-Bing Chen, Hui Gao, Linyuan L{\"u}, and Tao Zhou.
\newblock Identifying influential nodes in large-scale directed networks: the
  role of clustering.
\newblock {\em PloS One}, 8(10):e77455, 2013.

\bibitem{pastor2002epidemic}
Romualdo Pastor-Satorras and Alessandro Vespignani.
\newblock Epidemic dynamics in finite size scale-free networks.
\newblock {\em Physical Review. E}, 65(3):035108, 2002.

\bibitem{bogua2003epidemic}
Mari{\'a}n Bogu{\'a}, Romualdo Pastor-Satorras, and Alessandro Vespignani.
\newblock Epidemic spreading in complex networks with degree correlations.
\newblock {\em Statistical Mechanics of Complex Networks}, pages 127--147,
  2003.

\bibitem{barthelemy2005dynamical}
Marc Barth{\'e}lemy, Alain Barrat, Romualdo Pastor-Satorras, and Alessandro
  Vespignani.
\newblock Dynamical patterns of epidemic outbreaks in complex heterogeneous
  networks.
\newblock {\em Journal of Theoretical Bbiology}, 235(2):275--288, 2005.

\bibitem{centola2007complex}
Damon Centola and Michael Macy.
\newblock Complex contagions and the weakness of long ties.
\newblock {\em American Journal of Sociology}, 113(3):702--734, 2007.

\bibitem{centola2010spread}
Damon Centola.
\newblock The spread of behavior in an online social network experiment.
\newblock {\em Science}, 329(5996):1194--1197, 2010.

\bibitem{karsai2014complex}
M{\'a}rton Karsai, Gerardo Iniguez, Kimmo Kaski, and J{\'a}nos Kert{\'e}sz.
\newblock Complex contagion process in spreading of online innovation.
\newblock {\em Journal of the Royal Society Interface}, 11(101):20140694, 2014.

\bibitem{bianconi2021higher}
Ginestra Bianconi.
\newblock {\em Higher-order networks}.
\newblock Cambridge University Press, 2021.

\bibitem{alvarez2021evolutionary}
Unai Alvarez-Rodriguez, Federico Battiston, Guilherme~Ferraz de~Arruda, Yamir
  Moreno, Matja{\v{z}} Perc, and Vito Latora.
\newblock Evolutionary dynamics of higher-order interactions in social
  networks.
\newblock {\em Nature Human Behaviour}, 5(5):586--595, 2021.

\bibitem{sanchez2019high}
Alicia Sanchez-Gorostiaga, Djordje Baji{\'c}, Melisa~L Osborne, Juan~F Poyatos,
  and Alvaro Sanchez.
\newblock High-order interactions distort the functional landscape of microbial
  consortia.
\newblock {\em PLoS Biology}, 17(12):e3000550, 2019.

\bibitem{sizemore2018cliques}
Ann~E Sizemore, Chad Giusti, Ari Kahn, Jean~M Vettel, Richard~F Betzel, and
  Danielle~S Bassett.
\newblock Cliques and cavities in the human connectome.
\newblock {\em Journal of Computational Neuroscience}, 44:115--145, 2018.

\bibitem{spanier1989algebraic}
Edwin~H Spanier.
\newblock {\em Algebraic topology}.
\newblock Springer Science \& Business Media, 1989.

\bibitem{iacopini2019simplicial}
Iacopo Iacopini, Giovanni Petri, Alain Barrat, and Vito Latora.
\newblock Simplicial models of social contagion.
\newblock {\em Nature Communications}, 10(1):1--9, 2019.

\bibitem{battiston2020networks}
Federico Battiston, Giulia Cencetti, Iacopo Iacopini, Vito Latora, Maxime
  Lucas, Alice Patania, Jean-Gabriel Young, and Giovanni Petri.
\newblock Networks beyond pairwise interactions: structure and dynamics.
\newblock {\em Physics Reports}, 874:1--92, 2020.

\bibitem{torres2021and}
Leo Torres, Ann~S Blevins, Danielle Bassett, and Tina Eliassi-Rad.
\newblock The why, how, and when of representations for complex systems.
\newblock {\em SIAM Review}, 63(3):435--485, 2021.

\bibitem{battiston2021physics}
Federico Battiston, Enrico Amico, Alain Barrat, Ginestra Bianconi, Guilherme
  Ferraz~de Arruda, Benedetta Franceschiello, Iacopo Iacopini, Sonia K{\'e}fi,
  Vito Latora, Yamir Moreno, et~al.
\newblock The physics of higher-order interactions in complex systems.
\newblock {\em Nature Physics}, 17(10):1093--1098, 2021.

\bibitem{shi2019totally}
Dinghua Shi, Linyuan L{\"u}, and Guanrong Chen.
\newblock Totally homogeneous networks.
\newblock {\em National Science Review}, 6(5):962--969, 2019.

\bibitem{reimann2017cliques}
Michael~W Reimann, Max Nolte, Martina Scolamiero, Katharine Turner, Rodrigo
  Perin, Giuseppe Chindemi, Pawe{\l} D{\l}otko, Ran Levi, Kathryn Hess, and
  Henry Markram.
\newblock Cliques of neurons bound into cavities provide a missing link between
  structure and function.
\newblock {\em Frontiers in Computational Neuroscience}, page~48, 2017.

\bibitem{shi2013searching}
Dinghua Shi, Guanrong Chen, Wilson Wang~Kit Thong, and Xiaoyong Yan.
\newblock Searching for optimal network topology with best possible
  synchronizability.
\newblock {\em IEEE Circuits and Systems Magazine}, 13(1):66--75, 2013.

\bibitem{torres2020simplicial}
Joaqu{\'\i}n~J Torres and Ginestra Bianconi.
\newblock Simplicial complexes: higher-order spectral dimension and dynamics.
\newblock {\em Journal of Physics: Complexity}, 1(1):015002, 2020.

\bibitem{schaub2020random}
Michael~T Schaub, Austin~R Benson, Paul Horn, Gabor Lippner, and Ali Jadbabaie.
\newblock Random walks on simplicial complexes and the normalized hodge
  1-laplacian.
\newblock {\em SIAM Review}, 62(2):353--391, 2020.

\bibitem{mukherjee2016random}
Sayan Mukherjee and John Steenbergen.
\newblock Random walks on simplicial complexes and harmonics.
\newblock {\em Random Structures \& Algorithms}, 49(2):379--405, 2016.

\bibitem{courtney2016generalized}
Owen~T Courtney and Ginestra Bianconi.
\newblock Generalized network structures: the configuration model and the
  canonical ensemble of simplicial complexes.
\newblock {\em Physical Review. E}, 93(6):062311, 2016.

\bibitem{kovalenko2021growing}
Kiriil Kovalenko, Irene Sendi{\~n}a-Nadal, Nagi Khalil, Alex Dainiak, Daniil
  Musatov, Andrei~M Raigorodskii, Karin Alfaro-Bittner, Baruch Barzel, and
  Stefano Boccaletti.
\newblock Growing scale-free simplices.
\newblock {\em Communications Physics}, 4(1):1--9, 2021.

\bibitem{bobrowski2022random}
Omer Bobrowski and Dmitri Krioukov.
\newblock {\em Random simplicial complexes: models and phenomena}, pages
  59--96.
\newblock Springer, 2022.

\bibitem{costa2016random}
Armindo Costa and Michael Farber.
\newblock Random simplicial complexes.
\newblock In {\em Configuration Spaces}, pages 129--153. Springer, 2016.

\bibitem{yen2021construction}
Tzu-Chi Yen.
\newblock Construction of simplicial complexes with prescribed degree-size
  sequences.
\newblock {\em Physical Review. E}, 104(4):L042303, 2021.

\bibitem{C.elegans}
Duncan Watts and Steven Strogatz.
\newblock Collective dynamics of small world networks.
\newblock {\em Nature}, 393:440--2, 07 1998.

\bibitem{wikivote}
Roger Guimerà, Leon Danon, Albert Diaz-Guilera, Francesc Giralt, and Alex
  Arenas.
\newblock Self-similar community structure in a network of human interactions.
\newblock {\em Physical Review. E}, 68:065103, 01 2004.

\bibitem{email}
Ryan~A. Rossi and Nesreen~K. Ahmed.
\newblock The network data repository with interactive graph analytics and
  visualization.
\newblock In {\em Proceedings of the Twenty-Ninth AAAI Conference on Artificial
  Intelligence}, pages 4292--4293, 2015.

\bibitem{metabolic}
Jan Schellenberger, Junyoung Park, Tom Conrad, and Bernhard Palsson.
\newblock Bigg: a biochemical genetic and genomic knowledgebase of large scale
  metabolic reconstructions.
\newblock {\em BMC Bioinformatics}, 11:213, 04 2010.

\bibitem{shi2021computing}
Dinghua Shi, Zhifeng Chen, Xiang Sun, Qinghua Chen, Chuang Ma, Yang Lou, and
  Guanrong Chen.
\newblock Computing cliques and cavities in networks.
\newblock {\em Communications Physics}, 4(1):1--7, 2021.

\bibitem{shu2015numerical}
Panpan Shu, Wei Wang, Ming Tang, and Younghae Do.
\newblock Numerical identification of epidemic thresholds for
  susceptible-infected-recovered model on finite-size networks.
\newblock {\em Chaos: An Interdisciplinary Journal of Nonlinear Science},
  25(6):063104, 2015.

\bibitem{boguna2002epidemic}
Mari{\'a}n Bogun{\'a} and Romualdo Pastor-Satorras.
\newblock Epidemic spreading in correlated complex networks.
\newblock {\em Physical Review E}, 66(4):047104, 2002.

\bibitem{grigoriev1997pinning}
RO~Grigoriev, MC~Cross, and HG~Schuster.
\newblock Pinning control of spatiotemporal chaos.
\newblock {\em Physical Review Letters}, 79(15):2795, 1997.

\bibitem{wang2002synchronization}
Xiaofan Wang and Guanrong Chen.
\newblock Synchronization in small-world dynamical networks.
\newblock {\em International Journal of Bifurcation and Chaos},
  12(01):187--192, 2002.

\bibitem{wang2002pinning}
Xiaofan Wang and Guanrong Chen.
\newblock Pinning control of scale-free dynamical networks.
\newblock {\em Physica A: Statistical Mechanics and its Applications},
  310(3-4):521--531, 2002.

\bibitem{li2004pinning}
Xiang Li, Xiaofan Wang, and Guanrong Chen.
\newblock Pinning a complex dynamical network to its equilibrium.
\newblock {\em IEEE Transactions on Circuits and Systems I: Regular Papers},
  51(10):2074--2087, 2004.

\bibitem{fan2021characterizing}
Tianlong Fan, Linyuan L{\"u}, Dinghua Shi, and Tao Zhou.
\newblock Characterizing cycle structure in complex networks.
\newblock {\em Communications Physics}, 4(1):272, 2021.

\bibitem{xu2020quantifying}
Xiao-Ke Xu, Ke-Ke Shang, and Jing Xiao.
\newblock Quantifying the effect of community structures for link prediction by
  constructing null models.
\newblock {\em IEEE Access}, 8:89269--89280, 2020.

\bibitem{hong2004factors}
H~Hong, Beom~Jun Kim, MY~Choi, and Hyunggyu Park.
\newblock Factors that predict better synchronizability on complex networks.
\newblock {\em Physical Review E}, 69(6):067105, 2004.

\bibitem{nishikawa2003heterogeneity}
Takashi Nishikawa, Adilson~E Motter, Ying-Cheng Lai, and Frank~C Hoppensteadt.
\newblock Heterogeneity in oscillator networks: Are smaller worlds easier to
  synchronize?
\newblock {\em Physical Review Letters}, 91(1):014101, 2003.

\bibitem{comellas2007synchronizability}
Francesc Comellas and Silvia Gago.
\newblock Synchronizability of complex networks.
\newblock {\em Journal of Physics A: Mathematical and Theoretical},
  40(17):4483, 2007.

\bibitem{Dolphin}
David Lusseau, Karsten Schneider, Oliver Boisseau, Patti Haase, Elisabeth
  Slooten, and Stephen Dawson.
\newblock The bottlenose dolphin community of doubtful sound features a large
  proportion of long-lasting associations.
\newblock {\em Behavioral Ecology and Sociobiology}, 54:396--405, 01 2003.

\bibitem{football}
Michelle Girvan and Mark Newman.
\newblock Community structure in social and biological networks.
\newblock {\em Proceedings of the National Academy of Sciences of the United
  States of America}, 99:7821--6, 07 2002.

\bibitem{zachary1977information}
Wayne~W Zachary.
\newblock An information flow model for conflict and fission in small groups.
\newblock {\em Journal of Anthropological Research}, 33(4):452--473, 1977.

\bibitem{knuth1993stanford}
Donald~Ervin Knuth.
\newblock {\em The Stanford GraphBase: a platform for combinatorial computing},
  volume~1.
\newblock AcM Press New York, 1993.

\bibitem{polbooks}
V.~Krebs.
\newblock The political books network.
\newblock http://www.orgnet.com.

\bibitem{michael1997modeling}
Judd~H Michael and Joseph~G Massey.
\newblock Modeling the communication network in a sawmill.
\newblock {\em Forest Products Journal}, 47(9):25, 1997.

\bibitem{newman2006finding}
Mark~EJ Newman.
\newblock Finding community structure in networks using the eigenvectors of
  matrices.
\newblock {\em Physical Review E}, 74(3):036104, 2006.

\bibitem{lfrbenchmark}
Andrea Lancichinetti, Santo Fortunato, and Filippo Radicchi.
\newblock Benchmark graphs for testing community detection algorithms.
\newblock {\em Physical Review. E}, 78:046110, 11 2008.

\bibitem{modularity}
Mark Newman and Michelle Girvan.
\newblock Finding and evaluating community structure in networks.
\newblock {\em Physical Review. E}, 69:026113, 03 2004.

\bibitem{louvain}
Vincent Blondel, Jean-Loup Guillaume, Renaud Lambiotte, and Etienne Lefebvre.
\newblock Fast unfolding of communities in large networks.
\newblock {\em Journal of Statistical Mechanics Theory and Experiment},
  2008:P10008, 04 2008.

\bibitem{holme2016temporal}
Petter Holme.
\newblock Temporal network structures controlling disease spreading.
\newblock {\em Physical Review. E}, 94(2):022305, 2016.

\bibitem{ge2014epidemic}
Xin Ge, Li~Lili, and Hui Li.
\newblock Epidemic spreading and immunization on assortative degree mixing
  networks.
\newblock In {\em International Conference on Natural Computation}, pages
  1014--1019. IEEE, 2014.

\bibitem{zuev2015exponential}
Konstantin Zuev, Or~Eisenberg, and Dmitri Krioukov.
\newblock Exponential random simplicial complexes.
\newblock {\em Journal of Physics A: Mathematical and Theoretical},
  48(46):465002, 2015.

\end{thebibliography}

\end{document}